\documentclass[a4paper,fleqn]{cas-dc}

\usepackage[numbers]{natbib}
\usepackage{amssymb}
\usepackage{amsmath}
\usepackage{graphicx}
\usepackage{rotating}
\usepackage{color}
\usepackage{setspace}
\usepackage{mathdots}

\def\tsc#1{\csdef{#1}{\textsc{\lowercase{#1}}\xspace}}
\tsc{WGM}
\tsc{QE}
\tsc{EP}
\tsc{PMS}
\tsc{BEC}
\tsc{DE}

\begin{document}

\title[mode = title]{Altermagnetism without a long-range order}
\shorttitle{Altermagnetism without a long-range order}
\author[1,2]{V. E. Valiulin}[orcid=0000-0001-6643-6526]
\ead{val.valiulin@gmail.com}

\author[1,2]{A. V. Mikheyenkov}

\author[3,4]{K. I. Kugel}

\affiliation[1]{organization={Vereshchagin Institute of High Pressure Physics,
Russian Academy of Sciences},
                addressline={Moscow (Troitsk)},
                postcode={108840},
                country={Russia}}

\affiliation[2]{organization={Moscow Institute of Physics and Technology (National Research University)},
                addressline={Dolgoprudny},
                postcode={141701},
                country={Russia}}

\affiliation[3]{organization={Institute for Theoretical and Applied Electrodynamics
Russian Academy of Sciences},
                addressline={Moscow},
                postcode={125412},
                country={Russia}}

\affiliation[4]{organization={National Research University Higher School of Economics},
                addressline={Moscow},
                postcode={101000},
                country={Russia}}

\cortext[cor1]{Corresponding author: V. E. Valiulin}

\begin{abstract}
The Kugel--Khomskii spin--pseudospin model, originally developed for transition-metal compounds with orbital degrees of freedom, has recently been reinterpreted in the context of altermagnetism. In this work, we theoretically investigate the emergence of altermagnetic behavior in the absence of long-range magnetic or orbital order. Using the rotation-invariant Green's function method for the $SU(2) \times SU(2)$ symmetric model on a square lattice and on a linear chain, we analyze spin--spin and spin--pseudospin correlation functions, excitation spectra, heat capacity, and susceptibilities. We show that beyond a critical intersubsystem exchange $K_c(T)$, a composite state arises with nonzero spin--pseudospin correlations, even though the average spin and pseudospin at each site are zero. The excitation spectrum splits into acoustic and optical branches, with nodal lines along $q_x = q_y$ --- a direct signature of altermagnetic symmetry.
A peak in heat capacity and a jump in susceptibility are observed at the phase boundary.
In 1D, the phase boundary is nonmonotonic and demonstrates reentrant transition. These results establish the concept of an ``altermagnetic paramagnet'' or ``altermagnetic liquid'' without long-range order, relevant for low-dimensional and strongly fluctuating systems.
\end{abstract}


\maketitle

\section{Introduction}
\label{intro}

The spin--pseudospin model (the Kugel--Khomskii model), formulated more than half a century ago~\cite{Kugel73_SPJETP,Kugel82_SPU}, was initially associated with transition-metal compounds (see also a more recent review~\cite{Strelt17_PU}). In this case, the pseudospin corresponds to the doublet or triplet component of a five-fold degenerate $d$ level split by a crystal field.

After decades of near-inactivity, the model was revisited in several aspects at once. First, spin--pseudospin excitations, that is, in the language of the original model, orbital waves (orbitons), were theoretically investigated and experimentally detected~\cite{Brink98_PRB,Saitoh01_N,Brink01_PRL,Ishiha05_NJP,Schlap12_N,Wohlfe19_a,Li21_PRL,Martin24_PRL,Ji26_PRB}. Second, suggestions emerged regarding the possibility of a disordered state in the model, an orbital (spin--orbital) liquid, particularly in low-dimensional systems~\cite{Wang09_PRB,Zaanen98_a,Corboz12_PRX,Yamada21_PRB}. Furthermore, the model can be used to describe not only $d$-metal compounds, but also other systems with two interacting quantum subsystems (detailed analysis can be found in the recent work~\cite{Chen24_nQM}). In particular, one can mention the ``general-purpose quantum simulator'', namely, cold atoms in optical traps (see, e.g., \cite{Gorshk10_NP,Belemu17_PRB,Belemu18_NJP}).

Finally, as ideas of quantum entanglement penetrated condensed matter physics, it became apparent that the spin--pseudospin model provides an extremely convenient platform for studying many-body entanglement, which has spurred numerous works on this topic~\cite{Chen07_PRB,Brzezi11_PRB,Brzezi12_PRL,Brzezi12_JPCS,Oles12_JPCM,Lundgr12_PRB,You12_PRB,Oles13_a,Brzezi14_PRLa,You15_PRB,Man18_JPCM,Gotfry20_CM,Gotfry20_PRR,Valiul20_PRB,Pandey21_PRB,Mohapa22_JPCM,Valiul23_SPC}.


The areas listed above emerged and developed primarily in the current century. And in recent years, the interpretation of the spin-pseudospin model has undergone yet another reincarnation, triggered by the discovery of altermagnetism. Here, the pseudospin subsystem has a completely different nature. However, mathematically, the model remains the same. Thus, the Kugel--Khomskii model in its various forms has become a working tool for the theoretical study of altermagnetism~\cite{Camera25_n2MA,Ornell_26_PRB,Kausha26_a,Sichel25_a,Daghof25_a,Meier25_a}. The term ``Kugel--Khomskii spin-orbital altermagnet'' has even emerged. This allows, in the literal sense of the word, the results obtained for historically preceding realizations of the model to be transferred to the case of altermagnetism. Of course, a new interpretation of the input parameters and the results obtained is critically important.

This work considers a particular case of such a reinterpretation. We discuss the possibility of realizing the model's state in 2D and 1D with strong spin--pseudospin correlations, but without long-range order.

Note that the altermagnetism differs fundamentally from ferromagnetism and antiferromagnetism in that the net magnetization is zero, but the spin splitting of the bands depends on the momentum and the sign of the sublattice (``spin-polarized band structure without net magnetization''). In the Kugel--Khomskii model, this situation arises naturally if the pseudospin is responsible not only for orbital order, but for bond alternation or orbital polarization alternating across the sublattices. Therefore, a state with $\langle\textbf{S} \rangle = 0$ and $\langle\textbf{T} \rangle = 0$,  but with nonzero spin--pseudospin correlations, is a microscopic model of an antiferromagnetic liquid.

\section{Justification of the approach}
\label{approach}

\textit{In transition-metal oxides} and related materials, partially filled $d$ shells correspond to the presence of both spin and orbital degrees of freedom. The orbital sector can be described using pseudospins  $\mathbf{T_i}$ acting on a basis of, for example, two orbitals $e_g$ or orbitals $t_{2g}$, where the orbital order corresponds to a nonzero average value of $\mathbf{T_i}$. At the single-site level, the orbitals interact:

\noindent -- with the crystal field: the anisotropic local environment of $d$ elements splits the orbital multiplets;

\noindent -- directly with spin via exchange (superexchange, Hund's rule coupling) and, to a higher order, via the spin-orbit coupling (SOC). Furthermore, the overlap integrals $t_{ij}^{\alpha \beta}$ depend strongly on the specific type of orbitals and the direction of the bond.

At the same time, in  \textit{altermagnets}, the key microscopic components of non-relativistic spin splitting are

\noindent -- the alternation of spin-polarized wave functions on sublattices that are not equivalent in symmetry;

\noindent -- direction-dependent (orbitally selective) electronic hopping that lifts spin degeneracy.

Both of these phenomena are naturally linked to orbital physics: orbital order or orbitally selective hybridization differentiates sublattices and bonds, while the exchange interaction links spin to orbitals. For example, the crystal field first binds orbitals to specific sublattices (say, one sublattice favors $|3z^2-r^2\rangle$, and another --- $|x^2-y^2\rangle$ orbitals), then antiferromagnetic exchange fixes the spin on these orbitals, creating a  momentum-dependent ``spin-orbital locking''~\cite{Vila24_a} and, thus, an altermagnetic band structure. The resulting spin splitting is no longer a purely spin-dependent effect, but reflects a spin--orbital interaction that manifests itself, in particular, in optics.

In traditional antiferromagnets (e.g., RuO$_2$, MnTe), there is long-range magnetic order. However, in low dimensions (2D and 1D) or under strong fluctuations, this order is suppressed, but the symmetry of the exchange interaction remains antiferromagnetic. This means that the spectrum of elementary excitations (e.g., $\omega_{ac}(\textbf{q})$ and  $\omega_{opt}(\textbf{q})$) retains the splitting characteristic of antiferromagnets along certain directions in the Brillouin zone, even at $\langle\textbf{S} \rangle = 0$. Such a state can be called a ``fluctuating antiferromagnet'' or a ``short-range order antiferromagnetic paramagnet''.

We see that the defining symmetry of the altermagnetic system, that is, the relation $\varepsilon_\uparrow (\mathbf{k}) = \varepsilon_\downarrow (\hat{R}\mathbf{k})$ with  $\hat{R}$ being an operation of a non-primitive point group, stems from the spatial arrangement of bonds and orbitals, rather than exclusively in an ordered spin texture. This implies that a broad spectrum of phenomena bearing an imprint of antiferromagnetic symmetry can be preserved or even be dominant in the complete absence of  a long-range order (LRO).

In this sense, the rotation-invariant Green's function method (RGM), see details below, essentially corresponds to a description of ``altermagnetism without LRO'' regime. Recall that within this approach, the Mermin--Wagner theorem holds explicitly: all single-site averages vanish $\langle\widehat{{S}}_{\mathbf{i}}\rangle = \langle \widehat{{T}}_{\mathbf{i}}\rangle =0$, and at any nonzero temperature, there is no spin or orbital long-range order. Nevertheless, this demonstrates

\noindent -- the emergence, at the critical value of the spin-pseudospin interaction $K_c (T)$ ($T$ is the  temperature), of a composite state with spin-orbital entanglement, characterized by nonzero spin and pseudospin single-site $m_{0}=\langle S_{\mathbf{i}}^{z}T_{\mathbf{i}}^{z}\rangle$  and inter-site $m_{g}=\langle S_{\mathbf{i}}^{z}T_{\mathbf{i+g}}^{z}\rangle$ correlations ($\mathbf{i+g}$  are the nearest neighbors of site $i$). This strongly resembles, in fact, a second-order phase transition with spin--pseudospin correlator as an order parameter, a critical exponent $\alpha \sim 0.3-0.5$, and a phase boundary described in 2D by the relation  $T_c \approx 0.55 |K|^{0.55}$;

\noindent -- splitting of the collective excitation spectrum into the acoustic $\omega_{ac}(\mathbf{q})$ and the optical $\omega_{opt}(\mathbf{q})$ branches, defined by the parameters $K m_0$ and $\gamma_{\mathbf{q}}= \frac{1}{2} (\cos (q_{x})+\cos (q_{y}))$.

Parameter $\gamma_{\mathbf{q}}$ is identical to the $\mathbf{k}$-dependent structure factor that determines whether we have $\omega_{ac} \neq \omega_{opt}$ or not, whereas at the nodal lines $\cos (q_{x})+\cos (q_{y})$ (i.e., along the Brillouin zone diagonal $\mathbf{q} = (q,q)$) there appears a tendency to vanishing the spin gap, which is directly analogous to the nodal lines of the altermagnetic spin splitting along $\mathbf{k} = (k,k)$.

Note  that the analog to an order parameter here is not a spin or orbital expectation value, but some composite bilinear combination that locks spin and orbital, breaking SU(2)×SU(2) down to a residual SU(2) or U(1) in the combined space. That is qualitatively different from a spin–orbital liquid, where the full internal symmetry remains unbroken and only short-range entanglement exists.

An entangled state with $m_0 \neq 0$ does not violate any traditional spatial symmetry (all sites are equivalent, there is no sublattice magnetic moment), but it differentiates the spin and orbital sectors in a correlated, momentum-dependent manner. This is structurally analogous to an altermagnetic spin liquid: invariant with respect to traditional order parameters, but with nontrivial correlation functions.

As it was mentioned above, quantum entanglement itself in different various variants of the Kugel-Khomskii model has been previously studied. Here we only mention it, a detailed analysis in the context of altermagnetism is the subject of our further work.

Note also that the optical branch $\omega_{opt}(\mathbf{q})$ is analogous to an altermagnetic paramagnon mode: it is a propagating composite spin--orbital excitation that exists precisely because the spin--orbital entanglement corresponding to the short-range order lifts the spectrum degeneracy at $K=0$ without any type of condensation. In the language of the paramagnon--polaron model, this branch represents an ``altermagnon'' in the disordered spin--orbital liquid, which persists at temperatures above any ordering temperature, and causes a redistribution of the spectral weight of the $d$ wave characteristic of altermagnets.

\section{Model and method}
\label{model}

In this section, we will proceed from a symmetric version of the Kugel--Khomskii model, that is the $SU(2) \times SU(2)$ model with $SU(2)$ symmetries for both spin-$1/2$ and pseudospin-$1/2$ operators ( $\hat{\mathbf{S}}$ and $\hat{\mathbf{T}}$):

\begin{equation}
\widehat{\mathbf{H}} = J\sum_{<\mathbf{i},\mathbf{j}>}
{\mathbf{S}}_{\mathbf{i}}{\mathbf{S}}_{\mathbf{j}}+
I\sum_{<\mathbf{i},\mathbf{j}>}{\mathbf{T}}_{\mathbf{i}}
{\mathbf{T}}_{\mathbf{j}}
+K\sum_{<\mathbf{i},\mathbf{j}>}
\left({\mathbf{S}}_{\mathbf{i}}{\mathbf{S}}_{\mathbf{j}}\right)
\left( {\mathbf{T}}_{\mathbf{i}}{\mathbf{T }}_{\mathbf{j}}\right),
\label{Hamilt}
\end{equation}
where $<\mathbf{i},\mathbf{j}>$  denotes summation over the nearest-neighbor sites on a square lattice or linear chain.

We will limit ourselves to the most interesting case of antiferromagnetic (AFM) spin--spin and pseudospin--pseudospin interactions, assuming them to be equal, $J=I>0$. We presume the exchange coupling between subsystems to be negative $K<0$ (hereinafter, all energy parameters are given in the $J=I=1$ units).
Even at an early stage of the studies of this model, it was shown that at such a ratio of parameters, the inter-subsystem interaction is most pronounced~\cite{Pati98_PRL} and, in particular, can lead to an entanglement of two degrees of freedom~\cite{You12_PRB,Lundgr12_PRB}.

A key assumption for the subsequent discussion in this section is based on the paradigm of the absence of long-range order in dimensions $D  = 1$ and  2 at a temperature $T \neq 0$.  The reason for this choice is as follows. For non-interacting subsystems ($K = 0$), the Mermin--Wagner theorem~ \cite{Mermin66_PRL} prohibits a long-range order for  $D < 3$. It is natural to suppose that nonzero inter-subsystem interaction only enhances the role of thermal and quantum fluctuations.

In this section, we consider linear chains and square lattices and, as stated, assume that neither lattice nor spin symmetry is broken. There are several related theoretical approaches that satisfy these  conditions (see, e.g., the review article~\cite{Baraba26_PU}). Qualitatively, they are equivalent, and quantitatively, they yield similar results. Below, we adopt one of them, namely, the aforementioned  rotation-invariant Green's function method, RGM (see, e.g., \cite{Hartel11_PRB,Baraba11_TMP,Sherma12_JSNM,Mikhey13_JL,Hartel13_PRB,Mikhey16_JMMM,Valiul19_JPCM,Mueller19_PRB,Savche21_JMMM,Hutak23_EPJB}). Note that an alternative viewpoint, the emergence of long-range order in the system due to some mechanism, is also widely represented in works on spin models (for details, see in  \cite{Baraba26_PU}).

Thus, in this section, we proceed from \textit{the following assumptions}:

\textbf{i.} all sites are equivalent (translation symmetry is not broken);

\textbf{ii.} single-site averages are zero (no long-range order)
\begin{equation}
\langle\widehat{{S}}_{\mathbf{i}}\rangle =\langle \widehat{{T}}_
{\mathbf{i}}\rangle = 0; \label{site_aver1}
\end{equation}

\textbf{iii.} the correlation functions for the different spin and pseudospin components  ($\alpha \neq \beta$, $\alpha, \beta = x,y,z$) are also equal to zero ($SU(2)$  symmetries in spin and pseudospin spaces are not broken)
\begin{equation}
\langle \widehat{S}_{\mathbf{i}}^{\alpha}\widehat{S}_{\mathbf{j}}^{\beta}\rangle=0, \quad \langle \widehat{T}_{\mathbf{i}}^{\alpha}\widehat{T}_{\mathbf{j}}^{\beta}\rangle=0, \quad \langle \widehat{S}_{\mathbf{i}}^{\alpha}\widehat{T}_{\mathbf{j}}^{\beta}\rangle=0 . \label{ini_corrs}
\end{equation}

We consider the spin--spin and spin--pseudospin retarded Green's functions
\begin{equation}
G_{\mathbf{q}} = \left\langle S_{\mathbf{q}}^{z}\mid
S_{\mathbf{-q}}^{z}\right\rangle _{\omega } , \label{Gq01}
\end{equation}
\begin{equation}
R_{\mathbf{q}} = \left\langle T_{\mathbf{q}}^{z}\mid
S_{\mathbf{-q}}^{z}\right\rangle _{\omega } . \label{Rq01}
\end{equation}

For the symmetric case under study $I=J$, the third, pseudospin--pseudospin, Green's function is not needed, since $\left\langle T_{\mathbf{q}}^{z}\mid T_{\mathbf{-q}}^{z}\right\rangle _{\omega} = \left\langle S_{\mathbf{q}}^{z}\mid S_{\mathbf{-q}}^{z}\right\rangle_{\omega}$.

The calculations were performed following the standard RGB algorithm and using the approximation adopted in \cite{Kagan14_JL} for the $\widehat{K}$  term in  Hamiltonian (\ref{Hamilt}), which allows taking into account for inter-subsystem exchange. Then the expressions for $G_{q}^{zz}$ and $R_{\mathbf{q}}^{zz}$ take the form
\begin{equation}
G_{q} = \frac{F_{ac}(\mathbf{q})}{\omega ^{2}-\omega
_{ac}^{2}(\mathbf{q})}+\frac{F_{opt}(\mathbf{q})}{\omega ^{2}-\omega
_{opt}^{2}(\mathbf{q})} , \label{Gq1}
\end{equation}
\begin{equation}
R_{\mathbf{q}} = \frac{F_{ac}(\mathbf{q})}{\omega ^{2}-\omega
_{ac}^{2}(\mathbf{q})}-\frac{F_{opt}(\mathbf{q})}{\omega ^{2}-\omega
_{opt}^{2}(\mathbf{q})} , \label{Rq1}
\end{equation}

Rather cumbersome expressions for the numerators of the Green's functions and spectra of acoustic ($\omega_{ac}$) and optical ($\omega_{opt}$) excitation branches are given in the Appendix.

Expressions \eqref{Gq1} and \eqref{Rq1} include spin--spin and spin--pseudo-spin correlation functions. In the two-dimensional case, the spin--spin correlation functions are
\begin{equation}
c_{r}=\langle \hat{S}_{\mathbf{i}}^{z}\hat{S}_{\mathbf{i+r}}^{z}\rangle,
\quad r = g, d, 2g,
\end{equation}
respectively, for the first (side of the square) $c_{g} \equiv c_{1}$,  the second (diagonal) $c_{d} \equiv c_{2}$, and the third (doubled side) $c_{2g} \equiv c_{3}$ the nearest neighbors. In the one-dimensional case there are obviously no diagonal correlation functions.
Spin--pseudospin correlation functions, the on-site $m_{0}$ and the inter-site $m_{g} \equiv m_{1}$ are equal to
\begin{equation}
m_{0}=\langle S_{\mathbf{i}}^{z}T_{\mathbf{i}}^{z}\rangle,\ \
m_{g}=\langle S_{\mathbf{i}}^{z}T_{\mathbf{i+g}}^{z}\rangle ,
\end{equation}

Then all correlation functions involved into the problem, $c_{r}$ $(r=g, d, 2g)$ and $m_{0}$, $m_{g}$, are expressed in the standard way in terms of the Green's functions $G^{zz}$ and  $R^{zz}$. The resulting system of self-consistent equations is solved numerically. In this work, the case with $T \neq 0$ is considered.

\section{Results and discussion}
\label{results}

\subsection{Correlation functions in the 2$D$ case}

\begin{figure}
\centering
\includegraphics[width=.9\columnwidth]{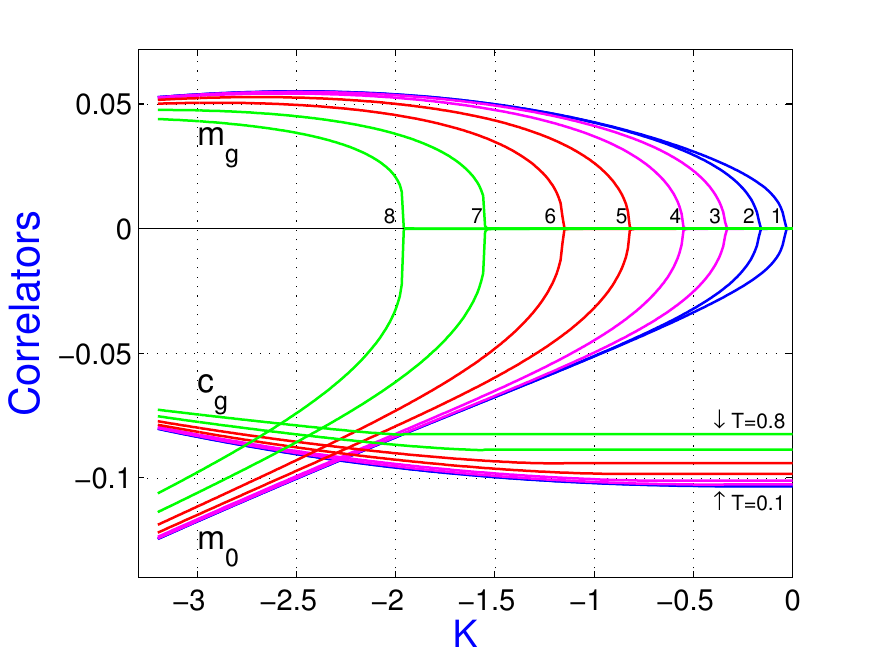}
\caption{\footnotesize (Color online) Dependence of the spin--spin correlation function $c_{g}$ at nearest neighbors and of the spin--pseudospin on-site $m_{0}$ and inter-site $m_{g}$ correlation functions on $T$ and on the inter-subsystem exchange $K$. The curves forming the ``platypus nose'' are $m_{0}$ ($m_{0}<0$) and $m_{g}$  ($m_{g}>0$). The numbers $1\div8$ enumerate the temperature values: 1 -- $T=0.1$, 2 -- $T=0.2$, etc. The lower curves correspond to $c_{g}$ (the limits of  $T$ are indicated). \label{corrs_on_KT}}
\end{figure}

The most notable manifestation of inter-subsystem interaction is the behavior of spin--pseudospin correlation functions. In Fig.~\ref{corrs_on_KT}, we illustrate for the square lattice the spin--spin correlation function $c_{g}$ for the nearest neighbors, as well as the spin--pseudospin correlation functions  $m_{0}$ and $m_{g}$ at the first and second coordination spheres, depending on temperature and inter-subsystem exchange parameter $K$.

For sufficiently small $K$,  the spin--pseudospin correlation functions, both single-site $m_{0}$, and inter-site $m_{g}$, vanish. However, at a certain critical value $K_c$, which depends on temperature, both correlation functions begin to increase steeply by absolute value ($m_{0}<0$, $m_{g}>0$). At the same time, the intra-subsystem (spin--spin) correlation function  varies smoothly, without any peculiarities. This provides, although not a strict proof, but a reasonable indication of the emergence of an ``entangled'' state in the system, characterized by nonzero spin--orbital correlation functions. The transition to this state resembles a second-order phase transition. Recall that in this case, a long-range order, both spin and orbital, is absent.

We see that at $|K| > |K_c|$, nonzero spin--pseudo-spin correlation functions ($m_{0}<0$ and $m_{g}> 0$) arise. In terms of altermagnetism, this corresponds to the emergence of a hidden spin--orbit correlation, which does not break the translational symmetry but causes an alternation in the sign of the spin polarization at different sublattices. A sign of $m_{0}<0$ indicates that the spin and pseudospin at a single site are anticorrelated, which is typical for an antiferromagnetic coupling between them. The sign $m_{g}> 0$ means that the spin at one site and the pseudospin at a neighboring site are positively correlated, this is a direct analogue of a ``long-range'' altermagnetic order in correlation functions without the net magnetization.

For a fixed value of the inter-subsystem exchange $K$, the temperature dependence of the spin--pseudospin correlation functions includes two regions. At temperatures above the critical one $T > T_c$, the correlation functions $m_{0}$ and $m_{g}$ are equal to zero, while at $T < T_c$, the behavior of both of them is well described by the power law $|m| \sim (T_c-T)^\alpha$ with an exponent $\alpha \sim 0.3\div0.5$ that depends weakly on $K$.

\subsection{2$D$ case: heat capacity and susceptibility}

\begin{figure}
\centering
\includegraphics[width=.9\columnwidth]{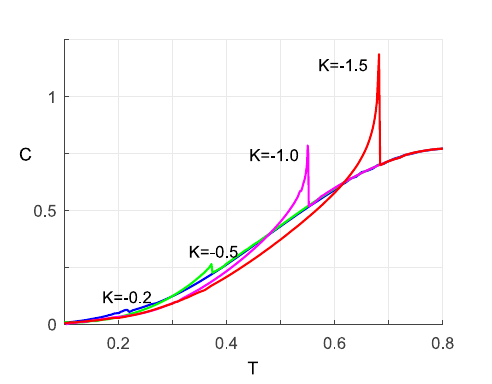}
\caption{\footnotesize (Color online) 2D lattice. Evolution of  the heat capacity with a change in $T$ at a fixed exchange parameter $K$. The colors correspond to different values of $K$ (values of $K$ are indicated near the peaks). Upon reaching the critical value $K_c$, a peak is observed, the height of which increases with $|K_c|$. \label{Fig_HC}}
\end{figure}

It is natural to expect that specific features of the correlation functions, the emergence of inter-subsystem correlations, entail the corresponding features of thermodynamic characteristics. The energy for the model under study is entirely determined by single-site and two-site correlation functions. Therefore, one could expect that a sharp increase in spin--pseudospin correlation functions in the entangled state will lead to a peculiarity in the heat capacity. In Fig.~\ref{Fig_HC}, we  show the temperature dependence of the heat capacity for several fixed $K$ values. As expected, upon entering the region with inter-subsystem correlations, the heat capacity undergoes a stepwise change. The higher the critical temperature $T_c$, the larger is the magnitude of this step. Fig.~\ref{Fig_HC} also shows that at high temperatures all curves converge to a single asymptote, and at $T \to 0$, we have $C \to 0$ for all curves (the Nernst theorem holds). Of course, as  $K$  changes at a fixed temperature, the heat capacity also undergoes a stepwise change at $K = K_c$.

\begin{figure}
\centering
\includegraphics[width=.9\columnwidth]{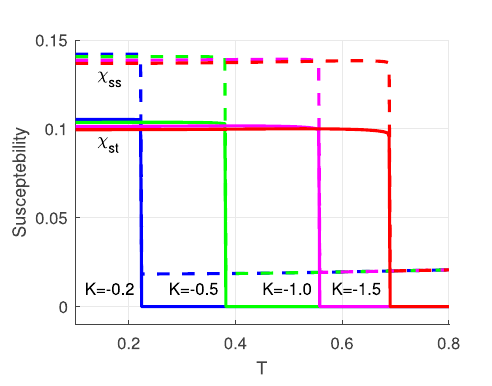}
\caption{\footnotesize (Color online) 2D lattice. Susceptibility as a function of temperature $T$ at the fixed inter-subsystem exchange $K$. The dashed line corresponds to the spin--spin susceptibility $\chi_{ss}$ and the solid line -- to the spin--pseudospin susceptibility $\chi_{st}$. Colors correspond to different values of $K$. \label{Fig_Susc}}
\end{figure}

Furthermore, Fig.~\ref{Fig_Susc} demonstrates  the temperature dependence of the spin--spin $\chi_{ss}$ and the spin--pseudospin $\chi_{st}$ susceptibilities at several fixed values of $K$. Similarly to the heat capacity, the susceptibility exhibits a stepwise change at critical points (the $K$ dependence $\chi_{ss}$ and $\chi_{st}$ at a fixed temperature looks similar to that of heat capacity). Outside the transition region, both susceptibilities depend only slightly on $T$ and $K$ . The behavior of the spin--spin susceptibility at small $|K|$ agrees with the well-studied case of the pure Heisenberg model (see, for example, the early work \cite{Shimah91_JPSJ}).

Thus, there is strong evidence that at any temperature with the growth of $K$ in the 2D case, that is, on a square lattice, there arises a state with entangled spin and pseudospin degrees of freedom. However, to rigorously confirm this conclusion, an accurate calculations of one of the entanglement measures are required.

\begin{figure}
\centering
\includegraphics[width=.9\columnwidth]{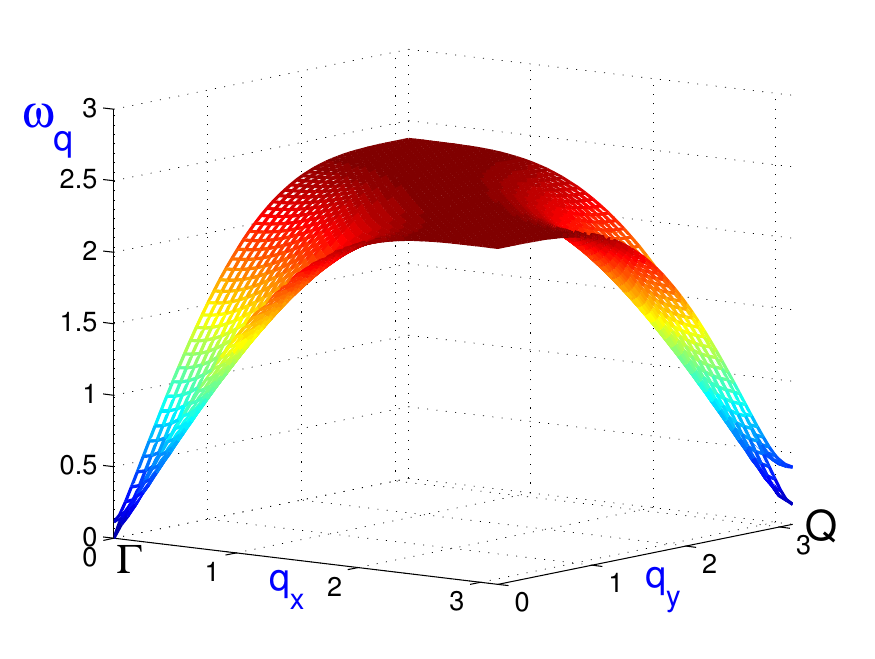}\\
\includegraphics[width=.9\columnwidth]{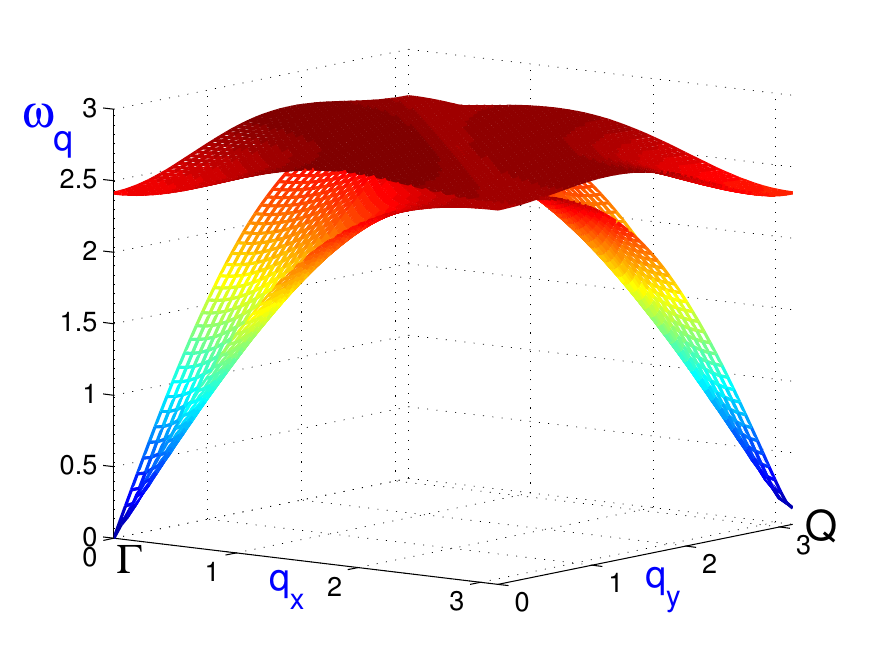}
\caption{\footnotesize (Color online) Spectra of elementary excitations $\omega_{ac}(\mathbf{q})$ and $\omega_{opt}(\mathbf{q})$ \eqref{spectra} at $T=0.3$. Top: $K=-0.4$, weak splitting. Bottom: $K=-3.0$, strong splitting. In the second case, the upper regions of the spectral branches form a nearly dispersionless region. A quarter of the Brillouin zone is shown. For any $T$ and $K$ at the symmetrical points $\mathbf{\Gamma}=(0,0)$ and $\mathbf{Q}=(\pi,\pi)$ in the Brillouin zone $\omega_{opt}(\mathbf{\Gamma})\geq\omega_{ac}(\mathbf{\Gamma})=0$ and $\omega_{ac}(\mathbf{Q})\geq\omega_{opt}(\mathbf{Q})\geq 0$. \label{Fig_spectra}}
\end{figure}

\begin{figure}
\centering
\includegraphics[width=.9\columnwidth]{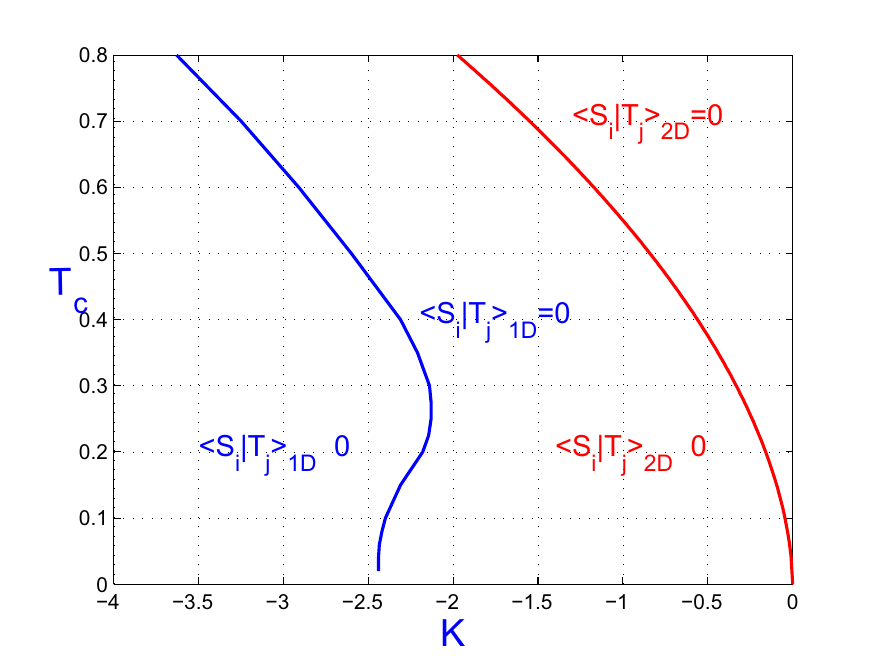}
\caption{\footnotesize (Color online)  Phase diagram - regions with zero and nonzero spin--pseudospin correlations are presented. The blue curve is the phase boundary in 1$D$ and the red curve -- in 2$D$. \label{Fig_PhaseDiag}}
\end{figure}

\subsection{2$D$ case: excitation spectrum}

Now let us turn to the spectrum of elementary excitations in the system.
When $K=0$ (and for any $|K| < |K_c|$), the subsystems $\mathbf{S}$ and $\mathbf{T}$ are independent  ($m_{0}=m_{g}=0$) and since we are dealing with the symmetric case $I=J$, their excitation spectra are identical. If $|K| > |K_c|$ is sufficiently large, the degenerate excitation spectrum splits into two branches $\omega_{ac}(\mathbf{q})$ and $\omega_{opt}(\mathbf{q})$. In the split case, for spectra at the high symmetry Brillouin zone points $\mathbf{\Gamma}=(0,0)$ and $\mathbf{Q}=(\pi,\pi)$, the following relations always hold
\begin{equation}
\omega_{opt}(\mathbf{\Gamma})\geq\omega_{ac}(\mathbf{\Gamma})=0, \quad
\omega_{ac}(\mathbf{Q})\geq\omega_{opt}(\mathbf{Q})\geq 0. \label{omegas}
\end{equation}
In particular, $\omega_{ac}$ is always a Goldstone branch.

In Fig.~\ref{Fig_spectra}, we illustrate the evolution of the elementary excitation spectrum with the growth of the inter-subsystem interaction $|K|$.  Near the transition, at $|K| \gtrsim |K_c|$, the spectral splitting is small and noticeable only in the vicinity of the highly symmetric points $\mathbf{\Gamma}=(0,0)$ and $\mathbf{Q}=(\pi,\pi)$ (see the upper panel in Fig.~\ref{Fig_spectra}). With an increase in $|K|$, the splitting grows. At a sufficiently large inter-subsystem exchange, the upper parts of the branches form a nearly dispersionless region (see the lower panel in Fig.~\ref{Fig_spectra}).

So for $|K| > |K_c|$, the spectrum splits into acoustic and optical branches, $\omega_{ac}(\textbf{q})$ and  $\omega_{opt}(\textbf{q})$. In altermagnetics, the splitting of spin bands is usually described as a $d$ wave $\sim(\cos{k_x} - \cos{k_y})$ or a $p$ wave.  Here, the analogue is $\gamma_{\textbf{q}} = (\cos{q_x} + \cos{q_y})$ (see the Appendix). It is particularly important that in the nodal directions ($q_x = q_y$) the splitting vanishes: this is a direct indication of altermagnetic symmetry in the excitation spectrum. The optical branch $\omega_{opt}(\textbf{q})$ is an ``altermagnon''(see recent~\cite{Liu24_PRL,Gauswa26_JPCM}),
that is a collective mode associated with fluctuations in the relative alignment of spin and pseudospin. Its existence above $T_c$ and without LRO is a key feature of an altermagnetic liquid.

The origin of the splitting is an exchange-driven composite order parameter $m_0$, that breaks the spin–orbital internal symmetry down to a diagonal subgroup and couples to bosonic modes with a form factor $\gamma_{\textbf{q}}$ fixed by the non-primitive point-group symmetry. The diagnostic that this is altermagnetic rather than generic hybridization is the presence of symmetry-protected nodal lines, a sharp onset at a phase transition driven by $m_0$, $m_g$, and the mapping of the two branches into each other under the altermagnetic operation $\hat{R}$, directly mirroring the $\varepsilon_\uparrow (\mathbf{k}) = \varepsilon_\downarrow (\hat{R}\mathbf{k})$
condition in electronic altermagnets.

\subsection{Phase diagram and the 1$D$ anomaly}

The calculations in the one-dimensional case are similar to those described above for the 2$D$ case (see the Appendix for the corresponding expressions). There are indications that, in the 1$D$ case, the used approach yields reasonable agreement for the thermodynamic properties with the results of the Bethe anzatz approximation and numerical data for finite Heisenberg chains~\cite{Suzuki94_JPSJ,Junger04_PRB,Hartel08_PRB,Valiul20_PRB,Valiul23_SPC}.

In Fig.~\ref{Fig_PhaseDiag}, we show the phase diagram for both dimensions demonstrating the regions with zero and nonzero spin--pseudospin correlations. In the 2D case, the phase boundary looks as is intuitively expected; it is well described by the relation $T_c=0.55|K|^{0.55}$.

The counterintuitive shape of the phase boundary in the 1D case is rather striking. Unlike the 2D case, in 1D, the curve separating the regions with zero and nonzero spin--pseudospin correlations begins at a nonzero inter-subsystem exchange, at the point $K\approx -2.44$,~ $T=0$. Furthermore, the boundary exhibits a nonmonotonic behavior, which implies the possibility of a reentrant transition to the entangled state. Nevertheless, as in 2D, the inter-subsystem correlation functions $m_{0}$ and $m_{g}$ are equal to zero until attaining the critical value $K_c$, and then they begin to increase steeply, whereas the change in the spin--spin correlation function is less noticeable. The heat capacity and susceptibility also exhibit peculiarities at the transition point. Due to the aforementioned nonmonotonity, the resulting plots, analogous to Figs.~\ref{corrs_on_KT}, \ref{Fig_HC}, and~\ref{Fig_Susc} have a more complicated shape, and we will not present them here.

In the context of altermagnetism, this is extremely interesting. A 1D chain is a model system for an altermagnetic spin liquid, where the long-range order is always suppressed, but the spin--pseudospin entanglement can emerge and vanish as $T$ changes. The nonmonotonity of the transition means that an increase in temperature can induce an entangled (altermagnetic) state being an effect reminiscent of entropy-stabilized altermagnetic phases. This prediction can be tested in cold atoms or organic chains.

A justification for this behavior in a one-dimensional chain can be obtained through an exact numerical analysis of inter-subsystem entanglement.

The reentrant behavior of entanglement in low-dimensional or small-particle spin models (including the Kugel-Khomskii one) has been repeatedly obtained by an exact diagonalization~\cite{Dajka08_PRA,Seidel19_PRL,Wang19_PRA,Ghanna21_M}.
A qualitative explanation is based on the fact that, at sufficiently low temperatures, states with lower energy are not necessarily more entangled (correlated) than those with a slightly higher energy.

\section{Conclusions}

We have demonstrated that the manifestations of altermagnetism do not necessarily require a long-range order. It is sufficient for the alternating sign of spin--pseudospin correlations on bonds to be preserved, along with the corresponding splitting of the excitation spectrum.
Such a state, an ``altermagnetic paramagnet'' or ``altermagnetic liquid'', can exist in 2D and 1D cases at finite temperatures and manifest itself in the heat capacity, susceptibility, as well as in the neutron scattering (dissipative altermagnon). This opens the way to the search for altermagnetism in low-dimensional systems, where magnetic order vanishes but altermagnetic exchange symmetry is preserved.

\appendix
\section{Appendix}
\label{App}

Spin--spin ($G_{\mathbf{q}}$) and spin--pseudospin ($R_{\mathbf{q}}$) retarded Green's function are
\begin{equation}
G_{\mathbf{q}} = \left\langle S_{\mathbf{q}}^{z}\mid
S_{\mathbf{-q}}^{z}\right\rangle _{\omega } , \label{Gq0}
\end{equation}
\begin{equation}
R_{\mathbf{q}} = \left\langle T_{\mathbf{q}}^{z}\mid
S_{\mathbf{-q}}^{z}\right\rangle _{\omega } , \label{Rq0}
\end{equation}
For the considered symmetric case $I=J$ the pseudospin--pseudospin Green's function is needles, because
$\left\langle T_{\mathbf{q}}^{z}\mid T_{\mathbf{-q}}^{z}\right\rangle _{\omega} = \left\langle S_{\mathbf{q}}^{z}\mid S_{\mathbf{-q}}^{z}\right\rangle_{\omega}$.

The RGM approach leads to the following expressions for $G_{\mathbf{q}}$ and $R_{\mathbf{q}}$:
\begin{equation}
G_{q} = \frac{F_{ac}(\mathbf{q})}{\omega ^{2}-\omega
_{ac}^{2}(\mathbf{q})}+\frac{F_{opt}(\mathbf{q})}{\omega ^{2}-\omega
_{opt}^{2}(\mathbf{q})} , \label{Gq}
\end{equation}
\begin{equation}
R_{\mathbf{q}} = \frac{F_{ac}(\mathbf{q})}{\omega ^{2}-\omega
_{ac}^{2}(\mathbf{q})}-\frac{F_{opt}(\mathbf{q})}{\omega ^{2}-\omega
_{opt}^{2}(\mathbf{q})} , \label{Rq}
\end{equation}

\vspace*{6pt}
\textbf{2D case: a square lattice}\\
The numerators for acoustic and optical branches are
\begin{equation}
F_{ac}=\frac{F_{1}+F_{2}}{2},\ F_{opt}=\frac{F_{1}-F_{2}}{2},
\end{equation}
\begin{equation}
F_{1}=-8Jc_{g}(1-\gamma _{\mathbf{q}})-Mm_{0}, \ F_{2}=Mm_{0}, \label{F12_2D}
\end{equation}
and the excitations spectra
\begin{equation}
\omega_{ac}^{2}(\mathbf{q})= W_1 + W_2,\ \omega _{opt}^{2}(\mathbf{q})= W_1 - W_2, \label{spectra}
\end{equation}
\begin{eqnarray}
W_1&=&2J^{2}(1\!-\!\gamma _{\mathbf{q}})\Bigl\{1
+4\bigl[\widetilde{c}_{2g}\!+\!2\widetilde{c}_{d}\!-\!\widetilde{c}_{g}(1\!+\!4\gamma _{\mathbf{q}})\bigr]\Bigr\}\!+\notag \\
&&+4JM(2\widetilde{m}_{g}-\widetilde{m}_{g}\gamma_{\mathbf{q}}-\widetilde{m}_{0}\gamma _{\mathbf{q}})+\frac{1}{8}M^{2}, \label{W_1_2D} \\
W_2&=&-4JM\left[ \widetilde{c}_{g}(1-\gamma _{\mathbf{q}})+\widetilde{m}_{g}-\widetilde{m}_{0} \gamma_{\mathbf{q}}\right]
 -\frac{1}{8}M^{2}, \label{W_2_2D}
\end{eqnarray}
here $c_{r}=\langle \widehat{S}_{\mathbf{i}}^{z}\widehat{S}_{\mathbf{i+r}}^{z}\rangle,
\quad r = g, d, 2g$ are spin--spin correlation functions, respectively for first (side of the square) $c_{g}$, second (diagonal) $c_{d}$ and third (doubled side) $c_{2g}$ nearest neighbors, $\widetilde{c}_{r} = \alpha_{r} {c}_{r}$ are correlation functions with vertex corrections (for more details see \cite{Baraba26_PU,Valiul20_PRB,Valiul23_SPC}). The lattice sum for square case $\gamma_{\mathbf{q}}=\frac{1}{4}\sum_{\mathbf{g}}e^{i\mathbf{qg}}=
\frac{1}{2}(\cos (q_{x})+\cos (q_{y}))$.

In equations \ref{F12_2D}--\ref{W_2_2D} $m_{0}$ and intersite $m_{g}$ are on-site and intersite spin--pseudospin correlation functions
\begin{equation}
m_{0}=\langle S_{\mathbf{i}}^{z}T_{\mathbf{i}}^{z}\rangle,\ \
m_{g}=\langle S_{\mathbf{i}}^{z}T_{\mathbf{i+g}}^{z}\rangle ,
\end{equation}
and for the inter-subsystem vertex corrections $\widetilde{m}_{0}=\alpha _{ST}^{0}m_{0}$, $\widetilde{m}_{g}=\alpha _{ST}^{g}m_{g}$ we adopted the approximation $\alpha _{ST}^{0}=\alpha _{ST}^{g}=1$\footnote{Extensive experience of RGM approach in various spin/pseudospin models (see Refs. in Sec.\ref{model}) shows that manipulating vertex corrections within reasonable limits does not lead to a qualitative change in the results, in the worst case --- only to a slight quantitative shift. The latter is immaterial for the purposes of this work.}. For simplicity, we use the notation $M=8Km_{0}$.

Note, that the following relations for the symmetrical points $\mathbf{\Gamma}=(0,0)$ and $\mathbf{Q}=(\pi,\pi)$ in the Brillouin zone are always fulfilled
\begin{equation}
\omega_{opt}(\mathbf{\Gamma})\geq\omega_{ac}(\mathbf{\Gamma})=0, \quad
\omega_{ac}(\mathbf{Q})\geq\omega_{opt}(\mathbf{Q})\geq 0. \label{omegas1}
\end{equation}

\vspace*{6pt}
\textbf{1D case: a linear chain}\\
\begin{equation}
F_{ac}=\frac{F_{1}+F_{2}}{2},\ F_{opt}=\frac{F_{1}-F_{2}}{2},
\end{equation}
\begin{equation}
F_{1}=-4Jc_{g}(1-\gamma _{\mathbf{q}})-Mm_{0}, \ F_{2}=Mm_{0} \label{F12_1D}
\end{equation}
\begin{eqnarray}
W_1&=&J^{2}(1-\gamma _{\mathbf{q}})\Bigl\{ 1+4\bigl[\widetilde{c}_{2g}-
\widetilde{c}_{g}(1+2\gamma _{\mathbf{q}})\bigr] \Bigr\} + \notag \\
&&2JM\left( 2\widetilde{m}_{g}-\widetilde{m}_{o}
\gamma_{\mathbf{q}}-\widetilde{m}_{g}\gamma _{\mathbf{q}}\right) +\frac{1}{8}M^{2}, \label{W_1} \\
W_2&=&-2JM\left[ \widetilde{c}_{g}(1-\gamma _{\mathbf{q}})+\widetilde{m}%
_{g}-\widetilde{m}_{0}\gamma _{\mathbf{q}}\right] -\frac{1}{8}M^{2}, \label{W_2}
\end{eqnarray}
now $M=4Km_{0}$, $\gamma_{\mathbf{q}}$ is one-dimensional, other notations are the same as for 2D case.

\section*{Acknowledgements}
The authors are grateful to V.V. Brazhkin and S.V. Streltsov for useful discussions.

\printcredits

\section*{Declaration of competing interest}
The authors declare that they have no known competing financial interests or personal relationships that could have appeared to influence the work reported in this paper.

\section*{Data availability}
No data was used for the research described in the article.

\end{document}